\documentstyle[12pt]{article}
\newcommand{\cl}{\centerline}
\renewcommand{\theequation}{\arabic{equation}}
\newcommand\beq{\begin{equation}}
\newcommand\eeq{\end{equation}}
\newcommand\bea{\begin{eqnarray}}
\newcommand\eea{\end{eqnarray}}

\begin{document}

\begin{titlepage}
\setlength{\textwidth}{5.0in}
\setlength{\textheight}{7.5in}
\setlength{\parskip}{0.0in}
\setlength{\baselineskip}{18.2pt}
\begin{center}
{\Large{{\bf BRST invariant $CP^{1}$ model through improved 
Dirac quantization}}}\par
\vskip 0.5cm
\begin{center}
{Soon-Tae Hong$^{1,2,a}$, Young-Jai Park$^{1,b}$, Kuniharu Kubodera$^{2,c}$}\par
{and Fred Myhrer$^{2,d}$}\par
\end{center}
\begin{center}
{$^{1}$Department of Physics and Basic Science Research Institute,}\par
{Sogang University, C.P.O. Box 1142, Seoul 100-611, Korea}\par
\vskip 0.2cm
{$^{2}$Department of Physics and Astronomy,}\par
{University of South Carolina, Columbia, SC 29208, USA}\par
\end{center}
\vskip 0.2cm
\cl{May 10, 2001}
\vfill
\begin{center}
{\bf ABSTRACT}
\end{center}
\begin{quotation}

The Batalin-Fradkin-Tyutin (BFT) scheme, which is an improved version of Dirac
quantization, is applied to the  $CP^1$ model, and the compact form of a nontrivial 
first-class Hamiltonian is directly obtained by introducing the BFT physical fields.  
We also derive a BRST-invariant gauge fixed Lagrangian through the standard 
path-integral procedure.  Furthermore, performing collective coordinate quantization 
we obtain energy spectrum of rigid rotator in the  $CP^1$ model.  Exploiting the Hopf 
bundle, we also show that the  $CP^1$ model is exactly equivalent to the O(3) nonlinear 
sigma model at the canonical level. 
   
\vskip 0.2cm
\noindent
PACS: 11.10.-z, 11.10.Ef, 11.30.-j, 14.20.-c\\
\noindent
Keywords: $CP^{1}$ model, Dirac quantization, BFT scheme, BRST symmetry\\
\noindent
---------------------------------------------------------------------\\
\noindent
$^a$sthong, $^b$yjpark@ccs.sogang.ac.kr; $^c$kubodera, $^d$myhrer@sc.edu
\vskip 0.5cm
\noindent
\end{quotation}
\end{center}
\end{titlepage}

\newpage

\section{Introduction}
\setcounter{equation}{0}
\renewcommand{\theequation}{\arabic{section}.\arabic{equation}}

Since the (2+1) dimensional O(3) nonlinear sigma model (NLSM) was first
discussed by Polyakov and Belavin~\cite{polyakov75}, there have
been lots of attempts to improve this soliton model associated
with the homotopy group $\pi_{2}(S^{2})=Z$.  The SU(N) invariant NLSM or 
$CP^{N-1}$ model was later introduced~\cite{eich78,witten79npb,adda79} in terms 
of $N$ complex fields $Z_{\alpha}$ ($\alpha=1,...,N$) satisfying a constraint 
$Z_{\alpha}^{*}Z_{\alpha}-1=0$.  In addition, one can impose a local 
U(1) invariance 
\beq
Z_{\alpha}(x)\rightarrow e^{i\alpha (x)}Z_{\alpha}(x)
\label{u1gauge}
\eeq
for arbitrary space-time dependent $\alpha (x)$~\cite{witten79npb}.  The 
$CP^{N-1}$ model for $N=2$ was shown~\cite{witten79npb} to be equivalent to 
the O(3) NLSM where Polyakov and Belavin found instantons~\cite{polyakov75}.  
Moreover, it is well known that the topological charge in the O(3) NLSM is 
equivalent to that in the SU(2) invariant NLSM~\cite{witten79npb}.

On the other hand, the Dirac method \cite{di} is a well known formalism 
to quantize physical systems with constraints.  In this method, the Poisson 
brackets in a second-class constraint system are converted into Dirac 
brackets to attain self-consistency.  The Dirac brackets, however, are 
generically field-dependent, nonlocal and contain problems related to ordering 
of field operators.  These features are unfavorable for finding canonically 
conjugate pairs. However, if a first-class constraint system can be constructed, 
one can avoid introducing the Dirac brackets and can instead use Poisson 
brackets to arrive at the corresponding quantum commutators.  

To overcome the above problems, Batalin, Fradkin, and Tyutin (BFT)~\cite{BFT} 
developed a method which converts the second-class constraints into 
first-class ones by introducing auxiliary fields. Recently, this BFT formalism 
has been applied to several models of current interest~\cite{BFT1,kpr,gafn}. In 
particular, the relation between the Dirac and BFT schemes, which had been 
obscure and unsettled, was clarified in the SU(2) Skyrmion model\cite{sk2}. 
Very recently, in Ref.~\cite{su2bft} the BFT Hamiltonian method was also applied to the SU(2) 
Skyrmion model to directly obtain the compact form of a first-class 
Hamiltonian, via the construction of the BFT physical fields.  Meanwhile, this 
BFT approach was applied to $CP^{1}$ model \cite{ban94, ban94r}, but these 
studies fell short of obtaining the desired compact form of a first-class 
Hamiltonian and as a result further developments have been deterred.

The motivation of this paper is to systematically apply the standard Dirac 
quantization method, the BFT scheme~\cite{kpr}, the Batalin, Fradkin and 
Vilkovisky (BFV) method~\cite{bfv,fik,biz} and the Becci-Rouet-Stora-Tyutin 
(BRST) method~\cite{brst} to the $CP^{1}$ model~\cite{ban94, ban94r}. In 
section 2 we convert the second-class constraints into first-class ones 
according to the BFT method to construct first-class BFT physical fields and 
directly derive the compact expression of a first-class Hamiltonian in terms 
of these fields.  The existing approach, used in Ref.~\cite{sk} to derive a 
first-class Hamiltonian in the SU(2) Skyrmion model, involves an infinite 
iteration procedure.  Our approach avoids this.  We then investigate some 
properties of the Poisson brackets of these BFT physical fields to obtain the 
Dirac brackets in the limit of vanishing auxiliary fields. We construct in 
section 3 a BRST-invariant gauge fixed Lagrangian in the BFV scheme through 
the standard path-integral procedure.  Exploiting collective coordinates, in 
section 4 we perform a semiclassical quantization and in section 5 we 
explicitly show the equivalence between the $CP^{1}$ model and O(3) nonlinear 
sigma model (NLSM)~\cite{bowick86} at the canonical level, by using the 
Hopf bundle~\cite{witten79npb,adda79}.  In section 6 we show that the 
energy spectrum of rigid rotator in the $CP^{1}$ model obtained by the 
standard Dirac method with the suggestion of generalized momenta is consistent 
with that of the BFT scheme.


\section{First-class constraints and Hamiltonian}
\setcounter{equation}{0}
\renewcommand{\theequation}{\arabic{section}.\arabic{equation}}


In this section we apply the BFT scheme to the $CP^{1}$
model, which is a second-class constraint system.  We start with the 
$CP^{1}$ model Lagrangian of the form 
\begin{equation}
L=\int {\rm d}^{2}x\left[\partial_{\mu}Z_{\alpha}^{*}\partial^{\mu}Z_{\alpha}
-(Z_{\alpha}^{*}\partial_{\mu}Z_{\alpha})(Z_{\beta}\partial^{\mu}
Z_{\beta}^{*})\right]
\label{cplag}
\end{equation}
where $Z_{\alpha}=(Z_{1},Z_{2})$ is a multiplet of complex scalar fields with a
constraint 
\begin{equation}
\Omega_{1}=Z_{\alpha}^{*}Z_{\alpha} - 1\approx 0.  \label{c1}
\end{equation}
One notes here that, as discussed before, this model is invariant under a 
local U(1) gauge symmetry transformation (\ref{u1gauge}).  By performing the 
Legendre transformation, one can obtain the canonical Hamiltonian 
\begin{equation}
H_c=\int{\rm d}^{2}x\left[\Pi_{\alpha}^{*}\Pi_{\alpha} 
+ \partial_{i}Z_{\alpha}^{*}
\partial_{i}Z_{\alpha}-(Z_{\alpha}^{*}\partial_{i}Z_{\alpha})
(Z_{\beta}\partial_{i}Z_{\beta}^{*})\right]
 \label{hc}
\end{equation}
where $\Pi_{\alpha}$ and $\Pi_{\alpha}^{*}$ are the canonical momenta 
conjugate to the complex scalar fields $Z_{\alpha}$ and $Z_{\alpha}^{*}$, 
respectively, given by 
\begin{eqnarray}
\Pi_{\alpha}&=&\dot{Z}_{\alpha}^{*}-Z_{\alpha}^{*}Z_{\beta}\dot{Z}_{\beta}^{*}
\nonumber\\
\Pi_{\alpha}^{*}&=&\dot{Z}_{\alpha}-Z_{\alpha}Z_{\beta}^{*}\dot{Z}_{\beta}.
\label{cojm}
\end{eqnarray}
The time evolution of the constraint $\Omega_1$ yields an additional secondary 
constraint 
\begin{equation}
\Omega_{2}=Z_{\alpha}^{*}\Pi_{\alpha}^{*}+Z_{\alpha}\Pi_{\alpha}\approx 0  
\label{const22}
\end{equation}
and $\Omega_{1}$ and $\Omega_{2}$ form a second-class constraint algebra 
\begin{equation}
\Delta_{kk^{\prime}}(x,y)=\{\Omega_{k}(x),\Omega_{k^{\prime}}(y)\}
=2\epsilon^{kk^{\prime}}Z_{\alpha}^{*}Z_{\alpha}\delta(x-y)  \label{delta}
\end{equation}
with $\epsilon^{12}=-\epsilon^{21}=1$.

Following the BFT formalism \cite{BFT,BFT1,kpr} which systematically
converts the second class constraints into first class ones, we introduce
two auxiliary fields $\Phi^{i}$ according to the number of second class
constraints $\Omega_{i}$ with the Poisson brackets 
\begin{equation}
\{\Phi^{i}(x), \Phi^{j}(y)\}=\epsilon^{ij}\delta(x-y). 
\label{phii}
\end{equation}
The first class constraints $\tilde{\Omega}_{i}$ fulfilling the simplest closed algebra  
\begin{equation}
\{\tilde{\Omega}_{i}(x),\tilde{\Omega}_{j}(y)\}=0  
\label{strong}
\end{equation}
are then constructed as follows
\begin{equation}
\tilde{\Omega}_{i}(x)=\Omega_{i}(x)+\int {\rm d}^{2}y X_{ij}(x,y)\Phi^{j}(y)  
\label{xijphi}
\end{equation}
where the matrix $X_{ij}$ satisfies the relation 
\begin{equation}
\Delta_{ij}(x,y)+\int {\rm d}^{2}z X_{ik}(x,z)
\epsilon^{kl}X_{jl}(z,y)=0.  
\label{delx}
\end{equation}
The solution of Eq. (\ref{delx}) is for instance given as 
\begin{equation}
X_{ij}(x,y)=\left( 
\begin{array}{cc}
2 & 0 \\ 
0 & -Z_{\alpha}^{*}Z_{\alpha}
\end{array}
\right)\delta (x-y)  \label{xij}
\end{equation}
to yield the first class constraints with the redefinition of the two auxiliary fields $\Phi^{i}=(\theta,\pi_{\theta})$  
\begin{eqnarray}
\tilde{\Omega}_{1}&=&\Omega_{1}+2\theta=Z_{\alpha}^{*}Z_{\alpha}-1+2\theta,  
\nonumber \\
\tilde{\Omega}_{2}&=&\Omega_{2}-Z_{\alpha}^{*}Z_{\alpha}\pi_{\theta}
=Z_{\alpha}^{*}\Pi_{\alpha}^{*}+Z_{\alpha}\Pi_{\alpha}-Z_{\alpha}^{*}
Z_{\alpha}\pi_{\theta}.
\label{1stconst}
\end{eqnarray}
Here one notes that the physical fields $Z_{\alpha}$ are geometrically 
constrained to reside on the $S^{3}$ hypersphere with the modified norm 
$Z_{\alpha}^{*}Z_{\alpha}=1-2\theta (x)$.

Now, we consider the uniqueness of the first class constraints.  In fact, 
according to the Dirac terminology~\cite{di}, the first class constraints 
$\tilde{\Omega}_{i}$ are defined to satisfy the following Lie 
algebra\footnote{In the case of the nonvanishing $C_{ij}^{k}$, Eq. 
(\ref{delx}) is modified as $\Delta_{ij}(x,y)+\int {\rm d}^{2}z X_{ik}(x,z)
\epsilon^{kl}X_{jl}(z,y)=C_{ij}^{k}(x,y)\tilde{\Omega}_{k}(y)$.}
\beq
\{\tilde{\Omega}_{i},\tilde{\Omega}_{j}\}=C_{ij}^{k}\tilde{\Omega}_{k}.
\label{weak}
\eeq
Since the first class constraints $\tilde{\Omega}_{i}$ are "strongly zero" 
$\tilde{\Omega}_{i}=0$ to yield 
$\{\tilde{\Omega}_{i}(x),\tilde{\Omega}_{j}(y)\}|{\rm phy}\rangle=0$ from 
Eq. (\ref{weak}), one does not have any difficulties in construction of the 
quantum commutators and in quantization of the given physical system.  
In that sense, one has degrees of freedom in taking a set of the first class 
constraints, without any criterion.  For instance our set of the first class 
constraints (\ref{1stconst}) is a specific choice to satisfy the minimal Lie 
algebra (\ref{strong}) with $C_{ij}^{k}=0$.  Moreover even in this minimal 
case, we have an equivalent family of the first class constraints governed 
by SO(2) rotation group, under which the matrix $X_{ij}$ transforms as 
\beq
X\rightarrow X^{\prime}=RXR^{T}
\label{rxr}
\eeq
where $R$ is an orthogonal $2\times 2$ matrix satisfying the condition 
$RR^{T}=1$.  Since the matrix $\epsilon$ is invariant under the SO(2) rotation, 
namely, $R\epsilon R^{T}=\epsilon$, one can easily check that the above 
rotated $X^{\prime}$ also satisfy the relation (\ref{delx}) to yield the 
equivalent family of first class constraints $\tilde{\Omega}_{i}^{\prime}$ 
given by inserting the $X^{\prime}$ into Eq. (\ref{xijphi}).  For the more
important case of the uniqueness of the first class Hamiltonian, we will 
discuss later in details.

Next, we construct the first class BFT physical fields 
$\tilde{{\cal F}}=(\tilde{Z}_{\alpha},\tilde{\Pi}_{\alpha})$ corresponding to 
the original fields ${\cal F}=(Z_{\alpha},\Pi_{\alpha})$.  The 
$\tilde{{\cal F}}$'s, which reside in the extended phase space, are obtained 
as a power series in the auxiliary fields $(\theta,\pi_{\theta})$ by demanding 
that they are strongly involutive: 
$\{\tilde{\Omega}_{i},\tilde{{\cal F}}\}=0$.  After some algebra, we obtain the 
first class physical fields as 
\begin{eqnarray}
\tilde{Z}_{\alpha}&=&Z_{\alpha}
\left(\frac{Z^{*}_{\beta}Z_{\beta}+2\theta}{Z^{*}_{\beta}Z_{\beta}}
\right)^{1/2}, 
\nonumber \\
\tilde{\Pi}_{\alpha}&=&\left(\Pi_{\alpha}-\frac{1}{2}Z_{\alpha}^{*}
\pi_{\theta}\right)\left(\frac{Z^{*}_{\beta}Z_{\beta}}
{Z^{*}_{\beta}Z_{\beta}+2\theta}\right)^{1/2}  
\label{pitilde}
\end{eqnarray}

As discussed in Ref.~\cite{kpr}, any functional ${\cal K}(\tilde{{\cal F}})$ 
of the first class fields $\tilde{{\cal F}}$ is also first class, namely, 
$\tilde{{\cal K}}({\cal F};\Phi )={\cal K}(\tilde{{\cal F}})$.  Using the 
property, we construct a first-class Hamiltonian in terms of the above BFT 
physical variables.  The result is  
\begin{equation}
\tilde{H}=\int {\rm d}^{2}x\left[\tilde{\Pi}_{\alpha}^{*}\tilde{\Pi}_{\alpha} 
+(\partial_{i}\tilde{Z}_{\alpha}^{*})(\partial_{i}\tilde{Z}_{\alpha})
-(\tilde{Z}_{\alpha}^{*}\partial_{i}\tilde{Z}_{\alpha})(\tilde{Z}_{\beta}
\partial_{i}\tilde{Z}_{\beta}^{*})\right].
\label{htilde}
\end{equation}
We then directly rewrite this Hamiltonian in terms of the original as well as 
auxiliary fields\footnote{%
In deriving the first class Hamiltonian $\tilde{H}$ of Eq. (\ref{hct}), we
have used the conformal map condition, $Z_{\alpha}^{*}\partial_{i}Z_{\alpha}
+Z_{\alpha}\partial_{i}Z_{\alpha}^{*}=0$.} to obtain
\begin{eqnarray}
\tilde{H}&=&\int {\rm d}^{2}x~\left[(\Pi_{\alpha}^{*}-\frac{1}{2}Z_{\alpha}
\pi_{\theta})(\Pi_{\alpha}-\frac{1}{2}Z_{\alpha}^{*}\pi_{\theta})
\frac{Z_{\beta}^{*}Z_{\beta}}{Z_{\beta}^{*}Z_{\beta}+2\theta}\right. 
\nonumber \\
& &\left.+(\partial_{i}Z_{\alpha}^{*})(\partial_{i}Z_{\alpha})
\frac{Z_{\beta}^{*}Z_{\beta}+2\theta}{Z_{\beta}^{*}Z_{\beta}}
-(Z_{\alpha}^{*}\partial_{i}Z_{\alpha})(Z_{\beta}\partial_{i}Z_{\beta}^{*})
\left(\frac{Z_{\gamma}^{*}Z_{\gamma}+2 \theta}{Z_{\gamma}^{*}Z_{\gamma}}\right)
^{2}\right].
\nonumber\\
\label{hct}
\end{eqnarray}
We observe that the forms of the first two terms in this Hamiltonian are 
exactly the same as those of the O(3) NLSM~\cite{o3}. 

Here $\tilde{H}$ is strongly involutive with the first class constraints 
$\{\tilde{\Omega}_{i},\tilde{H}\}=0$.
A problem with $\tilde{H}$ in Eq. (\ref{hct}) is that it does not naturally 
generate the first-class Gauss law constraint from the time evolution of the 
constraint $\tilde{\Omega}_{1}$.  Therefore, by introducing an additional term 
proportional to the first class constraints 
$\tilde{\Omega}_{2}$ into $\tilde{H}$, we obtain an equivalent first class 
Hamiltonian 
\begin{equation}
\tilde{H}^{\prime}=\tilde{H}+\frac{1}{2}\int {\rm d}^{2}x \pi_{\theta}\tilde{\Omega}_{2}  
\label{hctp}
\end{equation}
which naturally generates the Gauss law constraint 
\beq
\{\tilde{\Omega}_{1},\tilde{H}^{\prime}\}=\tilde{\Omega}_{2},~\{\tilde{\Omega}_{2},\tilde{H}^{\prime}\}=0.
\eeq
One notes here that $\tilde{H}$ and $\tilde{H}^{\prime}$ act in the same way 
on physical states, which are annihilated by the first-class constraints. 
Similarly, the equations of motion for observables remain unaffected by the 
additional term in $\tilde{H}^{\prime}$.  Furthermore, in the limit 
$(\theta,\pi_{\theta})\rightarrow 0$, our first class system is exactly reduced to the original second class one.

Now it is appropriate to comment on the uniqueness in construction of the 
first class Hamiltonians.  Similar to the above discussions on the uniqueness 
of the first class constraints, one can have the degrees of freedom in 
construction of the first class Hamiltonian $\tilde{H}$ or 
$\tilde{H}^{\prime}$ where $\tilde{H}^{\prime}$ is equivalent to $\tilde{H}$ 
up to the additional term $\tilde{\Omega}_{2}$ which does not affect the vacuum 
structure as discussed above.  However, imposing the condition that in the 
limit $(\theta,\pi_{\theta})\rightarrow 0$ the first class system is exactly 
reduced to the original second class one, one can exploit the degrees of the 
freedom to uniquely fix the specific form of first class Hamiltonian 
$\tilde{H}^{\prime}$ in Eq. (\ref{hctp}) which fulfills the Gauss law 
constraint.
   



Next, we consider the Poisson brackets of the fields in the extended phase
space $\tilde{{\cal F}}$ and identify the Dirac brackets by taking the
vanishing limit of auxiliary fields. After some algebraic manipulation 
starting from Eq. (\ref{pitilde}), one can obtain the commutators 
\begin{eqnarray}
\{\tilde{Z}_{\alpha}(x),\tilde{Z}_{\beta}(y)\}&=&
\{\tilde{Z}_{\alpha}^{*}(x),\tilde{Z}_{\beta}(y)\}=0,  \nonumber \\
\{\tilde{Z}_{\alpha}(x),\tilde{\Pi}_{\beta}(y)\}&=&(\delta_{\alpha\beta}
-\frac{\tilde{Z}_{\alpha}\tilde{Z}_{\beta}^{*}}{2\tilde{Z}_{\gamma}^{*}
\tilde{Z}_{\gamma}})\delta(x-y),  \nonumber \\
\{\tilde{Z}_{\alpha}(x),\tilde{\Pi}_{\beta}^{*}(y)\}&=&
-\frac{\tilde{Z}_{\alpha}\tilde{Z}_{\beta}}{2\tilde{Z}_{\gamma}^{*}
\tilde{Z}_{\gamma}}\delta(x-y),  \nonumber \\
\{\tilde{\Pi}_{\alpha}(x),\tilde{\Pi}_{\beta}(y)\}&=&\frac{1}
{2\tilde{Z}_{\gamma}^{*}\tilde{Z}_{\gamma}}
(\tilde{\Pi}_{\alpha}\tilde{Z}_{\beta}^{*}-\tilde{Z}_{\alpha}^{*}
\tilde{\Pi}_{\beta})\delta (x-y),\nonumber\\  
\{\tilde{\Pi}_{\alpha}(x),\tilde{\Pi}_{\beta}^{*}(y)\}&=&\frac{1}
{2\tilde{Z}_{\gamma}^{*}\tilde{Z}_{\gamma}}
(\tilde{\Pi}_{\alpha}\tilde{Z}_{\beta}-\tilde{Z}_{\alpha}^{*}
\tilde{\Pi}_{\beta}^{*})\delta (x-y).  
\label{commst}
\end{eqnarray}
In the vanishing auxiliary field limit, the above Poisson brackets in the
extended phase space exactly reproduce the corresponding Dirac brackets in 
the previous works~\cite{ban94, ban94r} 
\begin{eqnarray}
\{\tilde{Z}_{\alpha}(x),\tilde{Z}_{\beta}(y)\}
_{(\theta,\pi_{\theta})=0}
&=&
\{Z_{\alpha}(x),Z_{\beta}(y)\}_{D},  
\nonumber\\
\{\tilde{Z}_{\alpha}^{*}(x),\tilde{Z}_{\beta}(y)\}
_{(\theta,\pi_{\theta})=0}
&=&
\{Z_{\alpha}^{*}(x),Z_{\beta}(y)\}_{D},  
\nonumber\\
\{\tilde{Z}_{\alpha}(x),\tilde{\Pi}_{\beta}(y)\}
_{(\theta,\pi_{\theta})=0}
&=&
\{Z_{\alpha}(x),\Pi_{\beta}(y)\}_{D},  
\nonumber \\
\{\tilde{Z}_{\alpha}(x),\tilde{\Pi}_{\beta}^{*}(y)\}
_{(\theta,\pi_{\theta})=0}
&=&
\{Z_{\alpha}(x),\Pi_{\beta}^{*}(y)\}_{D},  
\nonumber \\
\{\tilde{\Pi}_{\alpha}(x),\tilde{\Pi}_{\beta}(y)\}
_{(\theta,\pi_{\theta})=0}
&=&
\{\Pi_{\alpha}(x),\Pi_{\beta}(y)\}_{D},
\nonumber\\  
\{\tilde{\Pi}_{\alpha}(x),\tilde{\Pi}_{\beta}^{*}(y)\}
_{(\theta,\pi_{\theta})=0}
&=&
\{\Pi_{\alpha}(x),\Pi_{\beta}^{*}(y)\}_{D}
\label{commstd}
\end{eqnarray}
where 
\begin{equation}
\{A(x),B(y)\}_{D}=\{A(x),B(y)\}-\int d^2z d^2 z^{\prime}
\{A(x),\Omega_{k}(z)\}\Delta^{k k^{\prime}}
\{\Omega_{k^{\prime}}(z^{\prime}),B(y)\}
\end{equation}
with $\Delta^{k k^{\prime}}$ being the inverse of $\Delta_{k k^{\prime}}$ in
Eq. (\ref{delta}).  It is also noteworthy that the Poisson brackets of 
$\tilde{{\cal F}}$'s in Eq. (\ref{commst}) have exactly the same form as 
those of the Dirac brackets of the field ${\cal F}$. In other words, the 
functional $\tilde{{\cal K}}$ in $\tilde{{\cal K}}({\cal F};\Phi )={\cal K}(\tilde{{\cal F}})$ 
corresponds to the Dirac brackets $\{A,B\}|_{D}$ and hence $\tilde{{\cal K}}$ corresponding to 
$\{\tilde{A},\tilde{B}\}$ becomes 
\begin{equation}
\{\tilde{A},\tilde{B}\}=\{A,B\}_{D}|_{A\rightarrow \tilde{A},B\rightarrow 
\tilde{B}}.
\end{equation}
This kind of situation happens again when one considers the first-class
constraints (\ref{1stconst}). More precisely, these first-class constraints
in the extended phase space can be rewritten as 
\begin{eqnarray}
\tilde{\Omega}_{1}&=&\tilde{Z}_{\alpha}^{*}\tilde{Z}_{\alpha}-1,  \nonumber \\
\tilde{\Omega}_{2}&=&\tilde{Z}_{\alpha}^{*}\tilde{\Pi}_{\alpha}^{*}
+\tilde{Z}_{\alpha}\tilde{\Pi}_{\alpha},  
\label{oott}
\end{eqnarray}
which are form-invariant with respect to the second-class constraints (\ref
{c1}) and (\ref{const22}).  


\section{BRST symmetries}
\setcounter{equation}{0}
\renewcommand{\theequation}{\arabic{section}.\arabic{equation}}


In this section we introduce two canonical sets of ghosts and anti-ghosts 
together with auxiliary fields in the framework of the BFV formalism 
\cite{bfv,fik,biz}, which is applicable to theories with the first-class 
constraints: 
\[
({\cal C}^{i},\bar{{\cal P}}_{i}),~~({\cal P}^{i}, \bar{{\cal C}}_{i}),
~~(N^{i},B_{i}),~~~~(i=1,2) 
\]
which satisfy the super-Poisson algebra 
\[
\{{\cal C}^{i}(x),\bar{{\cal P}}_{j}(y)\}=\{{\cal P}^{i}(x), \bar{{\cal C}}%
_{j}(y)\}=\{N^{i}(x),B_{j}(y)\}=\delta_{j}^{i}\delta(x-y). 
\]
Here the super-Poisson bracket is defined as 
\[
\{A,B\}=\frac{\delta A}{\delta q}|_{r}\frac{\delta B}{\delta p}|_{l}
-(-1)^{\eta_{A}\eta_{B}}\frac{\delta B}{\delta q}|_{r}\frac{\delta A} {%
\delta p}|_{l}, 
\]
where $\eta_{A}$ denotes the number of fermions, called the ghost number, 
in $A$ and the subscript $r$ and $l$ denote right and left derivatives, 
respectively.

In the $CP^{1}$ model, the nilpotent BRST charge $Q$ and the BRST invariant minimal
Hamiltonian $H_{m}$ are given by 
\begin{eqnarray}
Q&=&\int {\rm d}^{2}x~({\cal C}^{i}\tilde{\Omega}_{i}+{\cal P}^{i}B_{i}), 
\nonumber \\
H_{m}&=&\tilde{H}^{\prime}-\int {\rm d}^{2}x~{\cal C}^{1}\bar{{\cal P}}%
_{2},
\end{eqnarray}
which satisfy the relations 
\begin{equation}
\{Q,H_{m}\}=0,~~Q^{2}=\{Q,Q\}=0.
\end{equation}
Our next task is to fix the gauge, which is crucial to identify the BFT 
auxiliary field $\theta$ with the Stueckelberg field.  The desired 
identification follows if one chooses the fermionic gauge fixing function 
$\Psi$ as
\begin{equation}
\Psi=\int {\rm d}^{2}x~(\bar{{\cal C}}_{i}\chi^{i}+\bar{{\cal P}}%
_{i}N^{i}),  
\end{equation}
with the unitary gauge 
\begin{equation}
\chi^{1}=\Omega_{1},~~~\chi^{2}=\Omega_{2}.
\end{equation}
Here note that the $\Psi$ satisfies the following identity
\begin{equation}
\{\{\Psi,Q\},Q\}=0.
\end{equation}

The effective quantum Lagrangian is then described as 
\begin{equation}
L_{eff}=\int {\rm d}^{2}x~(\Pi_{\alpha}^{*}\dot{Z}_{\alpha}^{*}
+\Pi_{\alpha}\dot{Z}_{\alpha}
+\pi_{\theta}\dot{\theta} +B_{2}%
\dot{N}^{2}+\bar{{\cal P}}_{i}\dot{{\cal C}}^{i}+\bar{{\cal C}}_{2} \dot{%
{\cal P}}^{2})-H_{tot}
\end{equation}
where $H_{tot}=H_{m}-\{Q,\Psi\}$ and the terms $\int {\rm d}^{2}x~(B_{1}\dot{N}^{1}+\bar{{\cal C}}_{1}
\dot{{\cal P}}^{1})=\{Q,\int{\rm d}^{2}x~\bar{{\cal C}}_{1} \dot{N}^{1}\}$ 
have been suppressed by replacing $\chi^{1}$ with $\chi^{1} +\dot{N}^{1}$.

Now we perform path integration over the fields $B_{1}$, $N^{1}$, $\bar{%
{\cal C}}_{1}$, ${\cal P}^{1}$, $\bar{{\cal P}}_{1}$ and ${\cal C}^{1}$, by
using the equations of motion.  This leads to the effective Lagrangian of the 
form 
\begin{eqnarray}
L_{eff}&=&\int{\rm d}^{2}x~\left[\Pi_{\alpha}^{*}\dot{Z}_{\alpha}^{*}
+\Pi_{\alpha}\dot{Z}_{\alpha}+\pi_{\theta}\dot{\theta}
+B\dot{N}+\bar{{\cal P}}\dot{{\cal C}}+\bar{{\cal C}}\dot{{\cal P}}\right. 
\nonumber \\
& &\left.-(\Pi_{\alpha}^{*}-\frac{1}{2}Z_{\alpha}\pi_{\theta})
(\Pi_{\alpha}-\frac{1}{2}Z_{\alpha}^{*}\pi_{\theta}) 
\frac{Z_{\gamma}^{*}Z_{\gamma}}{Z_{\gamma}^{*}Z_{\gamma}+2\theta}
\right.
\nonumber\\
& &\left.-(\partial_{i}Z_{\alpha}^{*})(\partial_{i}Z_{\alpha})
\frac{Z_{\gamma}^{*}Z_{\gamma}+2\theta}{Z_{\gamma}^{*}Z_{\gamma}}
+(Z_{\alpha}^{*}\partial_{i}Z_{\alpha})
(Z_{\beta}\partial_{i}Z_{\beta}^{*})
\left(\frac{Z_{\gamma}^{*}Z_{\gamma}+2\theta}{Z_{\gamma}^{*}Z_{\gamma}}\right)^{2}  
\right.
\nonumber\\
& &\left.-\frac{1}{2}\pi_{\theta}\tilde{\Omega}_{2}+2Z_{\alpha}^{*}Z_{\alpha}
\pi_{\theta}\bar{{\cal C}}{\cal C}+\tilde{\Omega}_{2}N+B\Omega_{2}+\bar{{\cal P}}{\cal P}\right]
\end{eqnarray}
with the redefinitions: $N\equiv N^{2}$, $B\equiv B_{2}$, $\bar{{\cal C}}\equiv 
\bar{{\cal C}}_{2}$, ${\cal C}\equiv {\cal C}^{2}$, $\bar{{\cal P}}\equiv 
\bar{{\cal P}}_{2}$, ${\cal P}\equiv {\cal P}_{2}$.

After performing the routine variation procedure and identifying 
$N=-B+\frac{\dot{\theta}}{(1-2\theta)}$ we arrive at the effective 
Lagrangian of the form 
\begin{eqnarray}
L_{eff}&=&\int{\rm d}^{2}x~\left[\frac{1}{(1-2\theta)}
(\partial_{\mu}Z_{\alpha}^{*})(\partial^{\mu}Z_{\alpha})
-(1-2\theta)^{2}(B+2\bar{{\cal C}}{\cal C})^{2}\right.
\nonumber\\
& &\left. -\frac{1}{(1-2\theta)^{2}}
(Z_{\alpha}^{*}\partial_{\mu}Z_{\alpha})(Z_{\beta}\partial^{\mu}Z_{\beta}^{*})
-\frac{1}{1-2\theta}\partial_{\mu}\theta\partial^{\mu}B 
+\partial_{\mu}\bar{{\cal C}}\partial^{\mu}{{\cal C}}\right]
\nonumber\\
\label{leffbrst}
\end{eqnarray}
which is invariant under the BRST-transformation 
\begin{eqnarray}
\delta_{B}Z_{\alpha}&=&\lambda Z_{\alpha}{\cal C},~~~ 
\delta_{B}\theta=-\lambda (1-2\theta){\cal C},  \nonumber \\
\delta_{B}\bar{{\cal C}}&=&-\lambda B,~~~ \delta_{B}{\cal C}=\delta_{B}B=0.
\end{eqnarray}


\section{Collective coordinate quantization}
\setcounter{equation}{0}
\renewcommand{\theequation}{\arabic{section}.\arabic{equation}}


In this section, we perform a semi-classical quantization of the unit topological charge 
$Q=1$ sector of the $CP^{1}$ model by exploiting the collective coordinates 
to consider physical aspects of the theory.  
 
As a first approximation to the quantum ground state we could quantize zero modes
responsible for classical degeneracy by introducing collective coordinates
as follows 
\begin{eqnarray}
Z_{1}&=&e^{-i(\alpha+\phi)/2}\cos \frac{F(r)}{2},  \nonumber \\
Z_{2}&=&e^{i(\alpha+\phi)/2}\sin \frac{F(r)}{2},  \nonumber \\
\label{conf}
\end{eqnarray}
where $(r,\phi)$ are the polar coordinates and $\alpha (t)$ is the
collective coordinates. Here, in order to ensure the case of $Q=1$, 
we have used the fact the profile function $F(r)$ satisfies the boundary conditions: 
$\lim_{r\rightarrow \infty}F(r)=\pi$ and $F(0)=0$.

Using the above soliton configuration, we obtain the unconstrained
Lagrangian of the form 
\begin{equation}
L=-E+\frac{1}{2}{\cal I}\dot{\alpha}^{2},  \label{originl}
\end{equation}
where the soliton static mass and the moment of inertia are given by 
\begin{eqnarray}
E &=&\frac{\pi}{2}\int_{0}^{\infty }{\rm d}rr\left[\left(\frac{{\rm d}F}{{\rm d}r}
\right)^{2}+\frac{\sin ^{2}F}{r^{2}}\right] ,  \nonumber \\
{\cal I} &=&\pi\int_{0}^{\infty }{\rm d}rr\sin ^{2}F.
\label{inertias}
\end{eqnarray}
Introducing the canonical momentum conjugate to the collective coordinate $%
\alpha $ 
\begin{equation}
p_{\alpha }={\cal I}\dot{\alpha},
\end{equation}
we then have the canonical Hamiltonian 
\begin{equation}
H=E+\frac{1}{2{\cal I}}p_{\alpha }^{2}.  \label{hcc}
\end{equation}

At this stage, one can associate the Hamiltonian (\ref{hcc}) with the
previous one (\ref{hc}), which was given by the canonical momenta $\pi^{a}$.
Given the soliton configuration (\ref{conf}) one can obtain the relation
between $\pi^{a}$ and $p_{\alpha}$ as follows 
\begin{equation}
\Pi^{*}_{\alpha}\Pi_{\alpha}=\frac{\sin^{2}F}{4{\cal I}^{2}}p_{\alpha}^{2}  \label{rel}
\end{equation}
to yield the integral 
\begin{equation}
\int {\rm d}^{2}x \Pi^{*}_{\alpha}\Pi_{\alpha}=\frac{1}{2{\cal I}}%
p_{\alpha}^{2}.  \label{integ}
\end{equation}
Since the spatial derivative term in (\ref{hc}) yields nothing but the
soliton energy $E$, one can easily see, together with the relation (\ref
{integ}), that the canonical Hamiltonian (\ref{hc}) is equivalent to the
other one (\ref{hcc}), as expected.

Now, let us define the angular momentum operator $J$ as follows 
\begin{equation}
J=\int {\rm d}^{2}x \epsilon_{ij}x^{i}T^{oj},  
\label{jj}
\end{equation}
where the symmetric energy-momentum tensor is given by 
\begin{eqnarray}
T^{\mu\nu}&=&\partial^{\mu}Z^{*}_{\alpha}\partial^{\nu}Z_{\alpha}
+\partial^{\mu}Z_{\alpha}\partial^{\nu}Z^{*}_{\alpha}
-(Z_{\alpha}\partial^{\mu}Z^{*}_{\alpha})(Z^{*}_{\beta}\partial^{\nu}Z_{\beta})
\nonumber\\
& &-(Z^{*}_{\alpha}\partial^{\mu}Z_{\alpha})(Z_{\beta}\partial^{\nu}Z^{*}_{\beta})
   -g^{\mu\nu}(\partial_{\sigma}Z^{*}_{\alpha})(\partial^{\sigma}Z_{\alpha})
   +g^{\mu\nu}(Z^{*}_{\alpha}\partial_{\sigma}Z_{\alpha})(Z_{\beta}\partial^{\sigma}Z^{*}_{\beta}).
\nonumber\\
\label{tt}
\end{eqnarray}

Then, substituting the configuration (\ref{conf}) into Eq. (\ref{tt}), we
obtain the angular momentum operator of the form 
\begin{equation}
J=-{\cal I}\dot{\alpha}=-p_{\alpha}=i\frac{\partial}{\partial \alpha}
\end{equation}
to yield the Hamiltonian of the form 
\begin{equation}
H=E+\frac{1}{2{\cal I}}J^{2}.  \label{jham}
\end{equation}
Here one notes that the above Hamiltonian can be interpreted as mass spectrum of a
rigid rotator in the $CP^{1}$ model.

Next, let us consider the zero modes in the extended phase space by
introducing the soliton configuration 
\begin{eqnarray}
Z_{1}&=&(1-2\theta)^{1/2}e^{-i(\alpha+\phi)/2}\cos \frac{F(r)}{2},  \nonumber \\
Z_{2}&=&(1-2\theta)^{1/2}e^{i(\alpha+\phi)/2}\sin \frac{F(r)}{2},  
\label{conf2}
\end{eqnarray}
which satisfy the first class constraint $Z^{*}_{\alpha}Z_{\alpha}=1-2\theta$ of Eq. (\ref
{1stconst}).
\footnote{%
Here one can easily see that the first class physical fields $\tilde{Z}_{\alpha}$ of Eq. (%
\ref{pitilde}) satisfy the corresponding first class constraint $\tilde{Z}^{*}_{\alpha}
\tilde{Z}_{\alpha}=1$ of Eq. (\ref{oott}).} 
In this configuration from Eqs.
(2.1) and (5.11) we then obtain 
\begin{equation}
L_{eff}=-E+\frac{1}{2}{\cal I}\dot{\alpha}^{2},
\end{equation}
which is remarkably the Lagrangian (\ref{originl}) given in the original
phase space. Consequently the quantization of zero modes in the extended phase 
space reproduces the same energy spectrum (\ref{jham}). This phenomenon originates 
from the fact that the collective coordinates $\alpha$ in the Lagrangian (\ref{originl})
are not affected by the constraints (\ref{c1}) and (\ref{1stconst}) for the
complex scalar fields $Z_{\alpha}$. Here one notes that in the SU(2) Skyrmion model
the collective coordinates themselves are constrained to yield the modified
energy spectrum\cite{sk2,sk} in contrast to the case of the $CP^{1}$ model.


\section{Connection to O(3) nonlinear sigma model}
\setcounter{equation}{0}
\renewcommand{\theequation}{\arabic{section}.\arabic{equation}}


In this section, we will demonstrate the equivalence of the $CP^{1}$ model and 
O(3) NLSM~\cite{bowick86,o3} at the canonical level.  In the O(3) NLSM, the 
dynamical physical fields $n^{a}$ are mappings from the spacetime manifold, 
which is assumed to be the direct product of a compact two-dimensional Riemann 
surface ${\sf M}^{2}$ and the time dimension $R^{1}$, to the two-sphere 
$S^{2}$, namely $n^{a}: {\sf M}^{2}\otimes R^{1}\rightarrow S^{2}$.  On the 
other hand, the dynamical physical fields of the $CP^{1}$ model are 
$Z_{\alpha}$ which map the spacetime manifold ${\sf M}^{2}\otimes R^{1}$ into 
$S^{3}$, namely $Z_{\alpha}: {\sf M}^{2}\otimes R^{1}\rightarrow S^{3}$.  Here 
one notes that $S^{3}$ is homeomorphic to SU(2) group manifold.  

Since the $CP^{1}$ model is invariant under a local U(1) gauge symmetry, which 
consists of a redefinition of the phase of $Z_{\alpha}$ as in Eq. 
(\ref{u1gauge}), the physical configuration space of the $CP^{1}$ model are 
the gauge orbits which form the coset $S^{3}/S^{1}=S^{2}=CP^{1}$.  In order to 
associate the physical fields of the $CP^{1}$ model with those of the O(3) 
NLSM, we exploit the projection from $S^{3}$ to $S^{2}$, namely the Hopf 
bundle~\cite{witten79npb,adda79}\footnote{Here one notes that in order 
to eliminate all the unphysical degrees of freedom one can also supply a gauge 
fixing condition such as the Coulomb gauge: 
$Z_{\alpha}^{*}\partial_{i}\partial_{i}Z_{\alpha}-Z_{\alpha}\partial_{i}
\partial_{i}Z_{\alpha}^{*}=0$.}
\beq
n^{a}=Z^{*}_{\alpha}\sigma^{a}Z_{\alpha}
\label{bundle1}
\eeq
with the Pauli matrices $\sigma^{a}$, so that we can see that the $CP^{1}$ model 
Lagrangian (\ref{cplag}) is equivalent to the O(3) NLSM~\cite{bowick86} 
\begin{equation}
L=\int {\rm d}^{2}x \left[ \frac{1}{4}(\partial_{\mu}n^{a})
(\partial^{\mu}n^{a})\right]
\end{equation}
where $n^{a}$ ($a$=1,2,3) is a multiplet of three real scalar field with a
constraint 
\begin{equation}
\Omega_{1}=n^{a}n^{a}-1\approx 0.  \label{co3}
\end{equation}
Here note that the topological charge $Q=Z$ sector of the O(3) NLSM is 
guaranteed by the homotopy group $\pi_{2}(S^{2})=Z$. 

Moreover the collective coordinates (\ref{conf}) of the $CP^{1}$ model can 
be consistently obtained via the Hopf bundle (\ref{bundle1}) from those of 
the well known O(3) NLSM 
\begin{eqnarray}
n^{1}&=&\cos (\alpha (t)+\phi )\sin F(r),  \nonumber \\
n^{2}&=&\sin (\alpha (t)+\phi )\sin F(r),  \nonumber \\
n^{3}&=&\cos F(r),
\label{confo3}
\end{eqnarray}
to yield the rigid rotator energy spectrum (\ref{jham}), which is exactly 
the same as that of the O(3) NLSM~\cite{o3} with the same 
soliton static mass $E$ and moment of inertia ${\cal I}$ defined 
in Eq. (\ref{inertias}).

Now one can introduce the other bundle for the canonical momenta~\cite{ban94r}
\beq
\pi^{a}=\frac{1}{2}(\Pi_{\alpha}\sigma^{a}Z_{\alpha}
+Z^{*}_{\alpha}\sigma^{a}\Pi^{*}_{\alpha})
\label{bundle2}
\eeq
to reproduce the following secondary constraint from the corresponding $CP^{1}$ 
model one (\ref{const22}) 
\begin{equation}
\Omega_{2}=n^{a}\pi^{a}\approx 0.  \label{co32}
\end{equation}
Exploiting the above bundles (\ref{bundle1}) and (\ref{bundle2}) one can easily show that
the canonical Hamiltonian (\ref{hc}) of the $CP^{1}$ model is reduced to 
that of the O(3) NLSM 
\begin{equation}
H_c=\int{\rm d}^{2}x\left(\pi^{a}\pi^{a} 
+\frac{1}{4}\partial_{i}n^{a}\partial_{i}n^{a}\right).
\label{hco3}
\end{equation}
Here one notes that in contrast to the Banerjee case~\cite{ban94r}, where the 
reduced Hamiltonian has an additional term proportional to a first class constraint,
we have obtained the exactly same Hamiltonian as shown in (\ref{hco3}).   

Similarly, introducing the bundles for the first class physical fields $\tilde{n}^{a}$ 
and $\tilde{\pi}^{a}$
\bea
\tilde{n}^{a}&=&\tilde{Z}^{*}_{\alpha}\sigma^{a}\tilde{Z}_{\alpha}
\nonumber\\
\tilde{\pi}^{a}&=&\frac{1}{2}(\tilde{\Pi}_{\alpha}\sigma^{a}\tilde{Z}_{\alpha}
+\tilde{Z}^{*}_{\alpha}\sigma^{a}\tilde{\Pi}^{*}_{\alpha})
\label{bundle2t}
\eea
one can also find the equivalence between the $CP^{1}$ model and the O(3) NLSM 
in the extended phase space at the classical level.


\section{Connection to consistent Dirac quantization}
\setcounter{equation}{0}
\renewcommand{\theequation}{\arabic{section}.\arabic{equation}}


Now we consider consistent connection of the quantization of $CP^{1}$ model in the improved Dirac 
scheme to that of the standard Dirac one, where one can obtain the quantum commutators via 
Eqs. (\ref{commst}) and (\ref{commstd}) 
\begin{eqnarray}
{[Z_{\alpha}(x),Z_{\beta}(y)]}&=&
{[Z_{\alpha}^{*}(x),Z_{\beta}(y)]}=0,  \nonumber \\
{[Z_{\alpha}(x),\Pi_{\beta}(y)]}&=&i\left(\delta_{\alpha\beta}
-\frac{Z_{\alpha}Z_{\beta}^{*}}{2Z_{\gamma}^{*}Z_{\gamma}}\right)\delta(x-y),  \nonumber \\
{[Z_{\alpha}(x),\Pi_{\beta}^{*}(y)]}&=&-\frac{i}{2Z_{\gamma}^{*}Z_{\gamma}}Z_{\alpha}Z_{\beta}
\delta(x-y),  \nonumber \\
{[\Pi_{\alpha}(x),\Pi_{\beta}(y)]}&=&\frac{i}
{2Z_{\gamma}^{*}Z_{\gamma}}\left(\Pi_{\alpha}Z_{\beta}^{*}-Z_{\alpha}^{*}
\Pi_{\beta}\right)\delta (x-y),\nonumber\\  
{[Pi_{\alpha}(x),\Pi_{\beta}^{*}(y)]}&=&\frac{i}
{2Z_{\gamma}^{*}Z_{\gamma}}\left(\Pi_{\alpha}Z_{\beta}
-Z_{\alpha}^{*}\Pi_{\beta}^{*}\right)\delta (x-y),
\nonumber\\ 
\label{commst2}
\end{eqnarray}
where the quantum operator for the canonical momenta are given as 
\bea
\Pi_{\alpha}&=&-i(\delta_{\alpha\beta}-\frac{Z_{\alpha}^{*}Z_{\beta}}{2Z_{\beta}^{*}Z_{\beta}})
\partial_{\beta}\nonumber\\
\Pi_{\alpha}^{*}&=&-i(\delta_{\alpha\beta}-\frac{Z_{\alpha}Z_{\beta}^{*}}{2Z_{\beta}^{*}Z_{\beta}})
\partial_{\beta}^{*}
\label{quanops}
\eea
with the short hands $\partial_{\alpha}=\frac{\partial}{\partial Z_{\alpha}}$ and $\partial_{\alpha}^{*}=\frac{\partial}{\partial Z_{\alpha}^{*}}$.

Now we observe that without loss of generality the generalized momenta $%
\Pi_{\alpha}$ fulfilling the structure of the commutators (\ref{commst2}) is of the 
form 
\bea
\Pi_{\alpha}^{c}&=&-i(\delta_{\alpha\beta}-\frac{Z_{\alpha}^{*}Z_{\beta}}{2Z_{\beta}^{*}Z_{\beta}})
\partial_{\beta}-\frac{ic Z_{\alpha}^{*}}{2Z_{\beta}^{*}Z_{\beta}}
\nonumber\\
\Pi_{\alpha}^{c*}&=&-i(\delta_{\alpha\beta}-\frac{Z_{\alpha}Z_{\beta}^{*}}{2Z_{\beta}^{*}Z_{\beta}})
\partial_{\beta}^{*}-\frac{ic Z_{\alpha}}{2Z_{\beta}^{*}Z_{\beta}}
\label{cterm}
\eea
with an arbitrary parameter $c$ to be fixed later. 

On the other hand, the energy spectrum of the rigid rotators in the $CP^{1}$ 
model can be obtained in the Weyl ordering scheme~\cite{weyl} where the
Hamiltonian (\ref{hc}) is modified into the symmetric form 
\begin{equation}
H_{N}=E+\int {\rm d}^{2}x \Pi_{\alpha}^{N*}\Pi_{\alpha}^{N}
\label{hamil2}
\end{equation}
where 
\bea
\Pi_{\alpha}^{N}&=&-\frac{i}{2}\left[(\delta_{\alpha\beta}-\frac{Z_{\alpha}^{*}Z_{\beta}}{2Z_{\beta}^{*}Z_{\beta}})
\partial_{\beta}
+\partial_{\beta}(\delta_{\alpha\beta}
-\frac{Z_{\alpha}^{*}Z_{\beta}}{2Z_{\beta}^{*}Z_{\beta}})
+\frac{c Z_{\alpha}^{*}}{Z_{\beta}^{*}Z_{\beta}}\right]
\nonumber\\
\Pi_{\alpha}^{N*}&=&-\frac{i}{2}\left[(\delta_{\alpha\beta}-\frac{Z_{\alpha}Z_{\beta}^{*}}{2Z_{\beta}^{*}Z_{\beta}})
\partial_{\beta}^{*}
+\partial_{\beta}^{*}(\delta_{\alpha\beta}
-\frac{Z_{\alpha}Z_{\beta}^{*}}{2Z_{\beta}^{*}Z_{\beta}})
+\frac{c Z_{\alpha}}{Z_{\beta}^{*}Z_{\beta}}\right].
\label{cterm2}
\eea
After some algebra, one can obtain the Weyl ordered $\Pi_{\alpha}^{N*}
\Pi_{\alpha}^{N}$ as follows 
$$
\Pi_{\alpha}^{N*}\Pi_{\alpha}^{N}=-\partial_{\alpha}^{*}\partial_{\alpha}
+\frac{3Z_{\alpha}^{*}Z_{\beta}}{4Z_{\gamma}^{*}Z_{\gamma}}\partial_{\alpha}^{*}\partial_{\beta}
+\frac{3}{4Z_{\gamma}^{*}Z_{\gamma}}Z_{\alpha}\partial_{\alpha}
-\frac{1}{16Z_{\gamma}^{*}Z_{\gamma}}(2c-1)(2c+3)
$$
to yield the modified quantum energy spectrum of the rigid rotators 
\begin{equation}
\langle H_{N}\rangle=E+\frac{1}{2{\cal I}}J^{2}-\int {\rm d}^{2}x 
\frac{(2c+3)(2c-1)}{16}.
\label{hwc}
\end{equation}

Now, in order for the Dirac bracket scheme to be consistent with the BFT
one, the adjustable parameter $c$ in Eq. (\ref{hwc}) should be fixed with the
values 
\begin{equation}
c=\frac{1}{2},~-\frac{3}{2}.
\end{equation}
Here one notes that these values for the parameter $c$ relate the Dirac 
bracket scheme with the BFT one to yield the desired quantization in the 
$CP^{1}$ model so that one can achieve the unification of these two 
formalisms.


\section{Conclusions and Discussions}


In summary, we have constructed first-class BFT physical fields and, in
terms of them we have obtained a first-class Hamiltonian, consistent
with the Hamiltonian with the original fields and auxiliary fields. The
Poisson brackets of the BFT physical fields are also constructed and these 
Poisson brackets are shown to reproduce 
the corresponding Dirac brackets in the limit of vanishing auxiliary fields.  
Subsequently, we have obtained, in the Batalin, Fradkin and Vilkovisky (BFV) 
scheme\cite{bfv,fik,biz}, a BRST-invariant gauge fixed Lagrangian including 
the (anti)ghost fields, and BRST transformation rules under which the 
effective Lagrangian is invariant.  On the other hand, by performing the 
semiclassical quantization with the collective coordinates, we have obtained 
the spectrum of rigid rotator, which is shown to be consistent with that 
obtained by the standard Dirac method with the introduction of generalized 
momenta.  Next, using the Hopf bundle~\cite{bergeron92}, we have shown that 
the $CP^{1}$ model is exactly equivalent to the O(3) NLSM.  Through further 
investigation it will be interesting to include the Chern-Simons or Hopf term 
in the $CP^{1}$ model since there are still subtle ambiguities in the 
literatures~\cite{kov89}. 

Now, it is appropriate to comment on the dynamical aspects of the 
$CP^{N-1}$ model which possesses a local U(1) gauge invariance.  As discussed 
in Eq. (\ref{bundle1}), one has traded in two degrees of freedom 
(three components of $n^{a}$ minus one constraint (\ref{co3})) for the three 
degrees of freedom (four components of $Z$ minus one constraint (\ref{c1})).  
In fact, the $Z_{\alpha}$ field possesses only two degrees of freedom since an 
overall phase transformation (\ref{u1gauge}) does not change $n^{a}$ and hence 
the Lagrangian (\ref{cplag}).  In order to check that the Lagrangian 
(\ref{cplag}) is invariant under the U(1) local gauge transformation, one can 
rewrite the Lagrangian as~\cite{witten79npb,adda79}
\beq
L=\int {\rm d}^{2}x\left[\frac{2}{g^{2}}(D_{\mu}Z_{\alpha})^{*}(D^{\mu}Z_{\alpha})
+\lambda(Z_{\alpha}^{*}Z_{\alpha} - 1)\right]
\label{cplag2}
\eeq
with the covariant derivative $D_{\mu}=\partial_{\mu}-iA_{\mu}$ and the auxiliary gauge 
field $A_{\mu}=-iZ_{\alpha}^{*}\partial_{\mu}Z_{\alpha}$.  Here we have explicitly 
included the primary constraint $\Omega_{1}$ in Eq. (\ref{co3}) together with the 
Lagrangian multiplier field $\lambda$, and the coupling constant $g^{2}$~\cite{witten79npb}.  
Under the U(1) local gauge transformation (\ref{u1gauge}), one can have
\beq
A_{\mu}\rightarrow A_{\mu}+\partial_{\mu}\alpha
\label{au1}
\eeq
under which the Lagrangian (\ref{cplag2}) is invariant.  At the classical 
level, the gauge field $A_{\mu}$ associated with the U(1) symmetry is a 
redundant one which can be eliminated by using the equations of 
motion~\cite{witten79npb}.  

On the other hand, it was shown that in a stationary phase approximation the 
expectation value $\langle\lambda\rangle$ does not vanish for $g^{2}$ large 
enough~\cite{witten79npb}.  In this case, one can replace the field 
$\lambda (x)$ in the Lagrangian (\ref{cplag2}) with a constant 
$\lambda=\langle\lambda\rangle$ to yield the effective Lagrangian of the 
disordered phase where $Z_{\alpha}$ is an effective field no longer subjected 
to a constraint on its magnitude~\cite{witten79npb,zee91}.  Thus, at the quantum level, 
the $Z_{\alpha}$ field acquires a mass and the spin-spin correlation function 
becomes short ranged.  Here we have effectively traded in the constraint for 
a mass.  On the other hand, the gauge field $A_{\mu}$ may also acquire the kinetic term 
to become dynamical at the quantum level~\cite{witten79npb}.  Moreover, if the 
coupling constant $g^{2}$ becomes larger than some critical values, the symmetric 
phase appears with massless vector boson pole.  Since the O(3) NLSM has no local 
U(1) symmetry and may become singular due to a divergence in a composite vector boson 
channel, the above equivalence between the $CP^{1}$ model and O(3) NLSM breaks 
down at the quantum level where one needs to take into account properly the 
dynamical gauge boson effects.  

In the BFT scheme, at the quantum level, a similar situation happens to yield the 
quantum effects and the corresponding breakdown of the equivalence between the 
$CP^{1}$ model and the O(3) NLSM.  Moreover, through further investigation, it 
will be interesting to study a new term $\lambda\theta$ in 
$\lambda\tilde{\Omega}_{1}$ associated with the first class constraint 
$\tilde{\Omega}_{1}$ in Eq. (\ref{1stconst}), which may play a role 
in quantum level phenomenology.
 
\vskip 1.0cm 
STH would like to thank the University of South Carolina for the warm 
hospitality during his visit.  STH and YJP would like to thank B.H. Lee 
for helpful discussions and acknowledge financial support in part 
from the Korean Ministry of Education, BK21 Project No. D-1099 and Grant 
No. 2000-2-11100-002-5 from the Basic Research Program of the Korea Science 
and Engineering Foundation.  The work of KK and FM are supported in part by 
NSF (USA), Grant No. 9900756 and No. INT-9730847.

\end{document}